\documentclass[useAMS,usenatbib,usegraphicx]{mn2e}

\usepackage{amsmath}

\title[A MST algorithm for source detection in $\gamma$-ray images]
   {A MST algorithm for source detection in $\gamma$-ray images}
   
\author[R. Campana et al.]
  {Riccardo~Campana,$^1$\thanks{E-mail addresses: \texttt{riccardo.campana@uniroma1.it} 
  and \texttt{enrico.massaro@uniroma1.it}}
  Enrico~Massaro,$^{1,2}$\footnotemark[1] 
  Dario~Gasparrini,$^{2,3,4}$
 Sara~Cutini,$^{2,3,4}$  \newauthor
 and Andrea~Tramacere$^{5}$\\ \\
$^1$ Department of Physics, University of Rome ``La Sapienza'', Piazzale A. 
     Moro 2, I-00185, Rome, Italy \\
$^2$ ASI Science Data Center (ASDC), c/o ESRIN, Via G. Galilei, I-00044, Frascati, Italy \\
$^3$ Department of Physics, University of Perugia, Via A. Pascoli, I-06123, 
     Perugia, Italy \\
$^4$ INAF personnell resident at ASDC under ASI contract I/024/05/1  \\
$^5$ Stanford Linear Accelerator Center, 2575 Sand Hill Road, Menlo Park, CA-94025, USA \\
}
      
\date{Accepted 2007 October 19. Received 2007 September 25; in original form 2007 June 05.}

\pagerange{\pageref{firstpage}--\pageref{lastpage}} \pubyear{2007}

\begin{document}
\label{firstpage}

\maketitle

\begin{abstract}
We developed a source detection algorithm based on the Minimal Spanning Tree (MST), 
that is a graph-theoretical method useful for finding clusters in a given set of points. 
This algorithm is applied to $\gamma$-ray bidimensional images where the points 
correspond to the arrival direction of  photons, and the possible sources are associated 
with the regions where they clusterize.  
Some filters to select these clusters and to reduce the spurious detections are 
introduced.
An empirical study of the statistical properties of MST on random fields is carried in 
order to derive some criteria to estimate the best filter values.
We introduce also two parameters useful to verify the goodness of candidate sources.
To show how the MST algorithm works in the practice,
we present an application to an EGRET observation of the Virgo field, 
at high galactic latitude and with a low and rather uniform background, 
in which several sources are detected.
\end{abstract}


\section{Introduction}
Telescopes for satellite-based high energy  $\gamma$-ray astronomy 
detect individual photons by means of the electron-positron pair 
that they generate through the detector.
From the pair trajectories it is possible to reconstruct the original direction of
the photon with an uncertainty that decreases with the energy, from a few degrees 
below 100 MeV to less than a degree above 1 GeV. 
This technique was applied to the past $\gamma$-ray observatories SAS-2 (Fichtel 
et al. 1975), COS-B (Bennett 1990) and EGRET-CGRO (Kanbach et~al. 1988; Thompson et~al. 
1993), all equipped with spark chambers.
Pair tracking is also used in the current AGILE mission (Tavani et al. 2006) and 
in the LAT telescope on board the next GLAST mission, both employing silicon 
microstrip detectors (Gehrels et al. 1999). 
The resulting product is an image where each photon is associated with a direction in 
the sky: discrete sources thus correspond to regions in which a number of photons
higher than those found in the surroundings are observed.
When the size of this region is consistent with the instrumental Point Spread Function 
the source is considered as point-like, otherwise it can be extended or a group of 
near sources. 

Various algorithms are applied to the detection of point-like or extended sources 
in $\gamma$-ray astronomy: the most extensively used one is based on the 
Maximum Likelihood (Mattox et al. 1996), whereas others based on Wavelet Transform 
analysis (Damiani et al. 1997), Optimal Filter (Sanz et al. 2001), Scale-Adaptive 
Filter (Herranz et al. 2002), etc., were variously applied to real and simulated 
data to study their performances.
Some of them are based on deconvolution techniques of the instrumental 
Point Spread Function (PSF).
Many methods work directly on the pixellated images, i.e. count or intensity maps.
Other methods search for clusters in the arrival directions of photon that, if 
statistically significant, are considered an indication of a source.

The approach considered by us is essentially a cluster search based on a \emph{minimal 
spanning tree} (MST) algorithm.
This technique has its root in graph theory, and highlights the \emph{topological} 
pattern of connectedness of the detected photons.
Given a graph $G(V, E)$, where V is the set of vertices (or \emph{nodes}) and E is 
the set of weighted \emph{edges} connecting them, a MST (Kruskal 1956; Prim 1957; 
Zahn 1971) is the tree (a subgraph of $G$ without closed circuits) that connects 
all the points with the minimum total weight, defined as the sum of the weight of 
each tree's edge.
In a data set consisting of points in a Cartesian frame of reference, we can consider 
them as the nodes of a graph, the edges being the lines joining the nodes, weighted 
by their length.

The MST method was originally proposed for $\gamma$-ray source detection by 
Di~Ges\`u and Sacco (1983), who investigated also the statistical properties
in uniform fields.
This work was developed by Di~Ges\`u and Maccarone (1986), and De~Biase et al. (1986) 
applied MST for detecting extended sources in EXOSAT X-ray images.
Other authors applied MST methods to the goal of finding galaxy clusters, both 
in 2 and 3-dimensional surveys and simulations (Barrow et al. 1985; Bhavsar \& 
Ling 1988a,b; Plionis et al. 1992; Krzevina \& Saslaw 1996, Doroshkevich et al. 
2001, 2004) and showed the capabilites of the method as a filament-finding algorithm.

In this paper we investigate the MST approach in $\gamma$-ray source detection,
and present a new study of its statistical properties and the definition of selection
criteria. 
We also introduce some parameters useful to classify the reliabilty of detected
clusters to be associated with source candidates.  
We would like to emphasize here that this method is not \emph{alternative} to other 
source detection algorithms, but it is \emph{complementary}, in the 
sense that it can give a list of possible candidate sources (identified via their 
photons' clusterization properties) that could be further investigated by other means.

This paper is structured as follows.
In Sect.~2 we describe our MST algorithm, and in Sect.~3 and 4 we investigate by means of
numerical simulations the statistical distributions of edge length and node number, 
and we introduce some criteria useful for the source detection with our method.
An example of application to an EGRET field is shown in Sect.~5, while in Sect.~6 
we summarize and discuss our results.

\section{The MST algorithm}

\begin{figure*}
\centering
\includegraphics[scale=0.6]{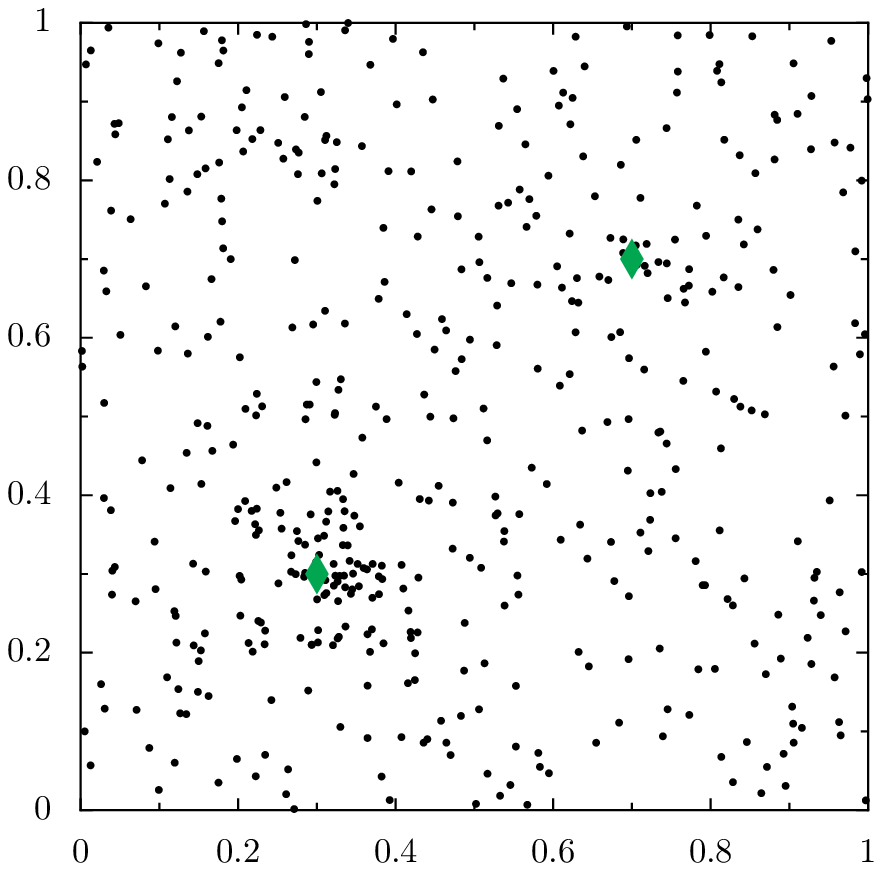} \includegraphics[scale=0.6]{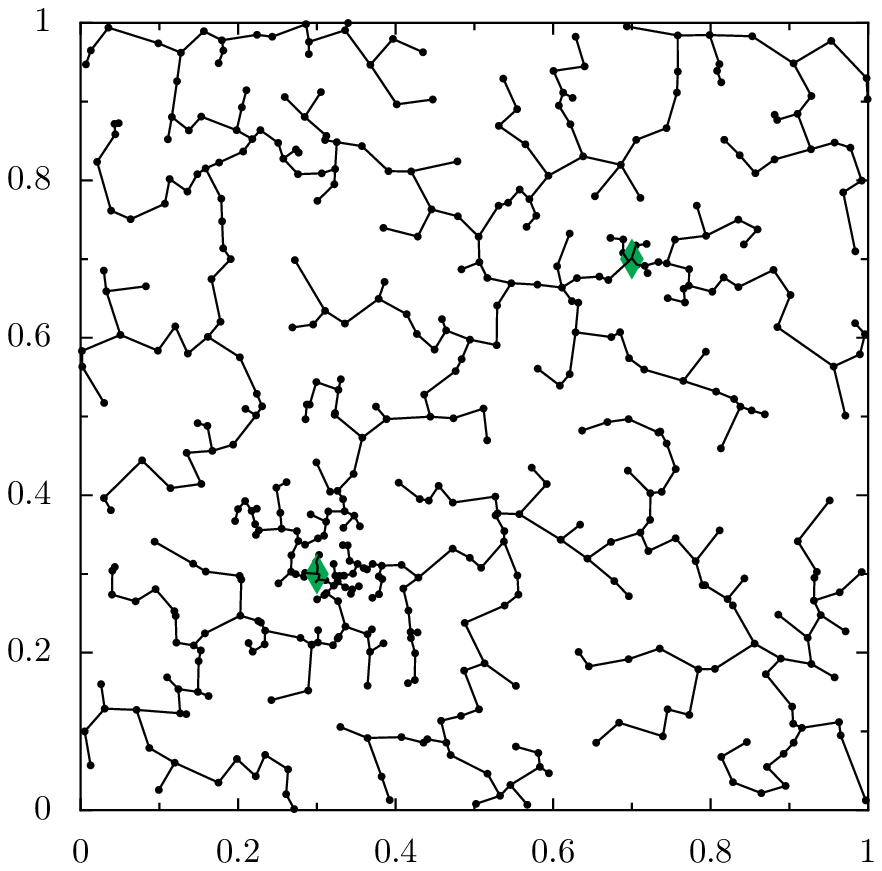} \\
\includegraphics[scale=0.6]{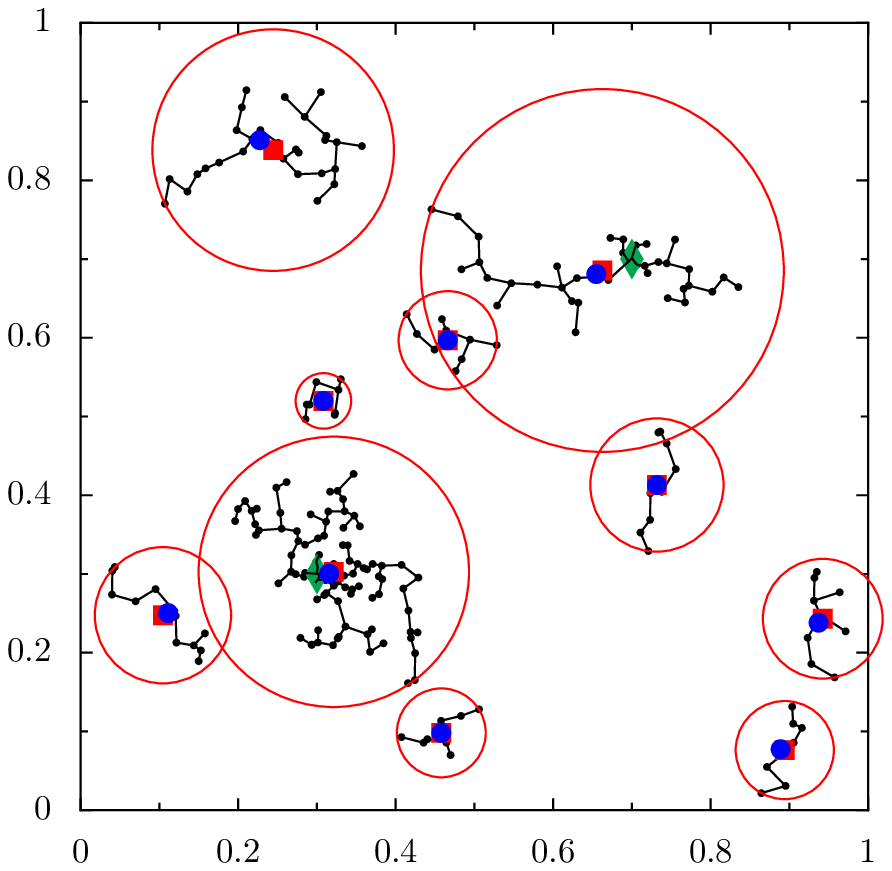} \includegraphics[scale=0.6]{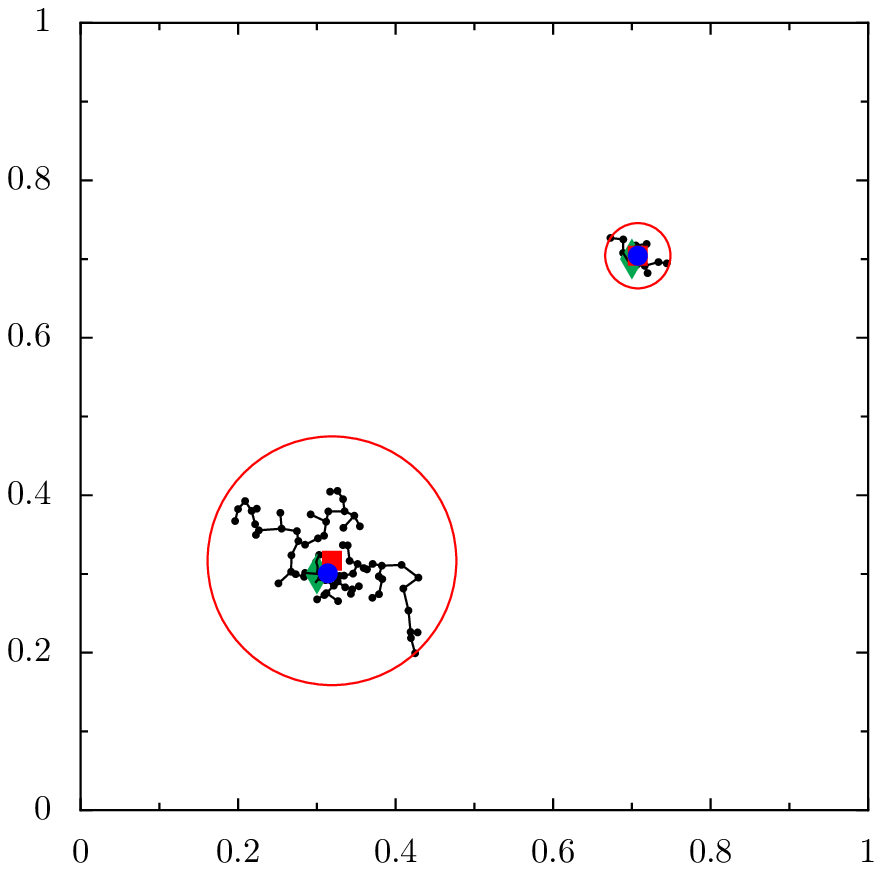}
\caption{\emph{Upper left}: A set of 500 random-generated points, with two simulated 
sources.
\emph{Upper right}: The Minimal Spanning Tree between these points.
\emph{Lower left}: Cluster selection after separation with $\Lambda_{\mathrm{c}}= 
1.3\,\Lambda_{\mathrm{m}}$ and elimination with $N_{\mathrm{c}}= 7$.
\emph{Lower right}: cluster selection with the filters 
$\Lambda_{\mathrm{c}}=\Lambda_{\mathrm{m}}$, $N_{\mathrm{c}}=10$.
The added ``sources'', at coordinates (0.3, 0.3) and (0.7, 0.7), are marked by the diamond. 
Circles are centered on the centroids of the remaining sub-trees (square) and 
have a radii equal to the distance of the farthest node in the sub-tree. 
The dot is the refined source position, see text for details.}
\label{fig:mstexample}
\end{figure*}

The result of an observation performed by a $\gamma$-ray telescope is a photon list 
containing for each event the arrival direction coordinates, time, energy, and other 
useful parameters.
Celestial coordinates (Right Ascension and Declination) of every photon define a point 
in a bi-dimensional frame and it can be considered a node in the graph. The edge weight
$\lambda$ is the angular distance between a couple of nodes.

The simplest way to find the MST of the field is a version of the Prim algorithm 
(also known as DJP algorithm; Prim 1957):
it starts from an arbitrary selected node, finds the nearest neighbour and connects 
them with an edge: this is the first edge of the MST. 
Then it finds the point that is the nearest to any point that is already connected 
in the MST. 
After $N-1$ iterations, where $N$ is the total number of points, the complete MST is 
found. 
Faster and computationally optimized algorithms can be found using other theoretical 
properties of the MST, like being a subset of the Delaunay triangulation of the graph 
(Delaunay 1934).
In particular, we used a fast code for the MST computation that is freely available 
from \textsc{Boost}\footnote{\texttt{http://www.boost.org}} and 
CGAL\footnote{\texttt{http://www.cgal.org}} 
libraries.

Once found the MST, to extract \emph{only} the locations where the photon 
clusterize, i.e. the possible sources, and to evaluate the residual photon background, 
the following operations must be performed:

\begin{itemize}
\item \textbf{Separation:} remove all the edges having $\lambda$ greater than a 
selected separation value $\Lambda_{\mathrm{c}}$. 
Usually, it is chosen in units of the mean edge length $\Lambda_{\mathrm{m}}$ in 
the MST. 
As a result we obtain a set of disconnected sub-trees.
\item \textbf{Elimination:} remove all the sub-trees having a number of nodes 
$N_\mathrm{n}$ less than or equal to a threshold value $N_{\mathrm{c}}$. 
This filter is useful to remove small casual clusters of nodes, leaving only the 
clusters 
that have a high probability to be genuine sources.
\end{itemize}

After the application of these filters, the remaining sub-trees correspond to 
possible sources.
An estimate of the source position is obtained by computing the centroid of the 
sub-tree nodes (i.e. the mean value of the Right Ascension and Declination between 
all points in the sub-tree). 
A refined source position can be found by computing the centroid of all the points 
lying inside the circle centered on the previous calculated sub-tree centroid with a 
radius equal to the distance of the farthest point in the sub-tree,
to take into account also possible photons belonging to the source but 
accidentally filtered out.

An example of this procedure is shown in Fig.~\ref{fig:mstexample}, where the 
upper-left panel shows a frame containing $N_{\mathrm{tot}}=500$ points within a 
square region of unit length: two clusters having different numbers of points 
have been added to a random generated point distribution. 
The first one, representative of a ``strong'' source, has 80 points spread on a 
Gaussian circle of $\sigma = 0.1$, the second one, the ``faint'' source, has 20 
points distributed in a similar circle. 
The random ``background'' has thus 400 points.
The upper-right panel shows the MST that connects all the points.
In the lower-left panel are shown the clusters detected after separation with 
$\Lambda_{\mathrm{c}}=1.3\, \Lambda_{\mathrm{m}}$ and elimination with 
$N_{\mathrm{c}}= 7$. 
In this case a few small size clusters are detected, which disappear when more 
appropiate filters are used ($\Lambda_{\mathrm{c}}=\Lambda_{\mathrm{m}}$, 
$N_{\mathrm{c}}=10$, lower-right panel), whereas the two genuine sources remain.
Their positions, computed from the sub-tree centroids, have a distance smaller
than 0.01 from the right ones, confirming the validity of this method to evaluate
the source coordinates.

The two major points of the MST source detection are therefore the choice of 
the two filtering parameters and the methods to evaluate the significance of the 
residual sub-trees.

\section{MST statistical properties}

\subsection{The length distribution in the MST}

\begin{figure}
\centering
\includegraphics[scale=0.4]{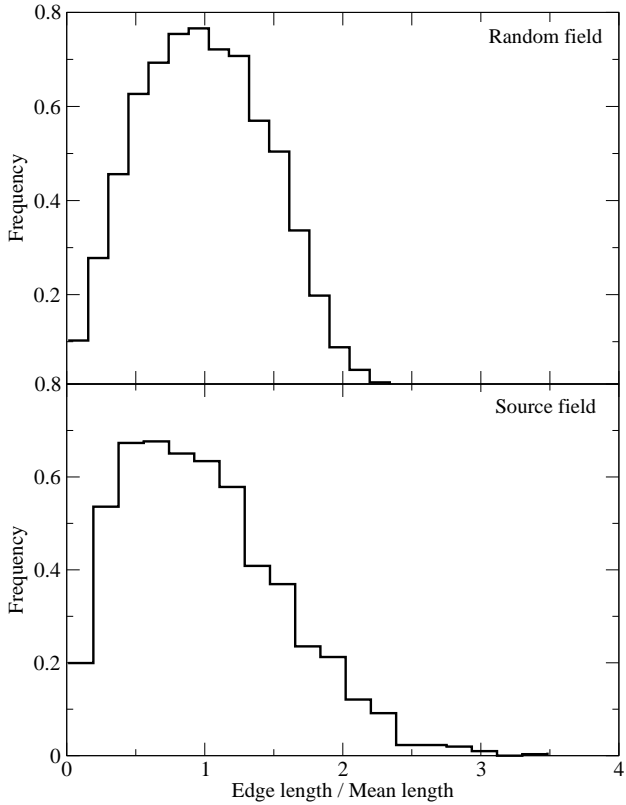}
\caption{\emph{Upper panel}: Histogram of the MST edge length, in units of the mean 
length, for a random field with 1675 points.
\emph{Lower panel}: Histogram of the MST edge length, in units of the mean length, 
for the same field in which some strong sources have been added. Note that there 
is a a large left-side asymmetry with respect to the random field.}
\label{fig:hist12}
\end{figure}

According to Barrow et al. (1985), an useful criterion to distinguish a random 
(Poissonian) field from a field with some sources, is the shape of the frequency 
distribution of the edge length in the MST. 
These authors suggest that for a random field this distribution has an approximate 
Gaussian shape peaked around the mean edge length $\Lambda_{\mathrm{m}}$.
We studied this distribution, whose statistical properties are useful to choose the
best filtering parameters.
In Fig. \ref{fig:hist12} we present the frequency distributions of 
$x = \lambda/\Lambda_{\mathrm{m}}$ computed for a frame with a random field 
(upper panel) and the same frame with five sources added (lower panel): 
in the latter case there is a clear excess of short distances (within the 
clusters that mark the sources) and of long distances (between the clusters) with 
respect to the random case, and the histogram shows an evident left asymmetry. 

A useful indicator for the presence of sources is the mean value of the MST length
$\Lambda_{\mathrm{m}}$. 
Earlier investigations (Gilbert 1965) found that the total length of a random MST is 
proportional to $\sqrt{(AN_{\mathrm{tot}})}$ where $A$ is the field area and 
$N_{\mathrm{tot}}$ is the total number of points. 
A theoretical upper limit to the proportionality constant was found to be 
$2^{-1/2} \simeq 0.70$. 
Our Monte Carlo simulations showed that the constant value is rather $\simeq 0.65$. 
Therefore the mean length for a random-field MST is:
\begin{equation}\label{e:meanlength}
\Lambda_{\mathrm{m}} \simeq 0.65 \times \sqrt{\frac{A}{N_{\mathrm{tot}}}}
\end{equation}
Thus, if the mean length for a field deviates from this value, it is an indicator of 
non-random clusterization, i.e. of the presence of sources.

Another test for the occurrence of sources is the evaluation of the skewness 
coefficient $\beta_3$ of the distribution $f(x)$.
In the two cases of Fig. \ref{fig:hist12} we found $\beta_3$ equal to 0.16 and 0.46;
the higher value is due to the decrease of the mean length $\Lambda_{\mathrm{m}}$ 
and to the occurrence of $x$ values greater than $\approx2.5$ when sources are present. 
From our simulations we found that $\beta_3$ higher than $\sim0.2$ can be considered 
a good indicator for the presence of sources.

For an accurate study of the edge length distribution it is useful to have a
simple analytical formula to be applied in the computation. 
Since  theoretical works on this subject are not easily available in the astronomical
literature, we followed a numerical approach. 

\begin{figure}
\centering
\includegraphics[scale=0.3, angle=270]{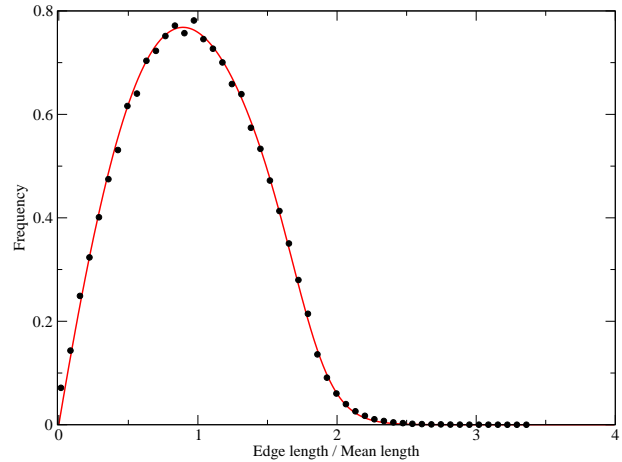}
\caption{Histogram of the MST edge length frequency, in units of the mean length, 
for a random field with $10^6$ points. 
Also plotted is Eq. \eqref{e:mstdistr3}.}\label{fig:hist3}
\end{figure}

First we generated a pure random frame containing $10^6$ points to smooth the
fluctuations in the histogram and the resulting frequency plot is given in 
Fig. \ref{fig:hist3}.
Note that, like in Fig. \ref{fig:hist12}, it has a well defined mode, a small
skewness and very small tail for $x > 2$. 
Its shape is not, therefore, that of a Gaussian and an approximate formula that 
gives an excellent best fit, although properly it is defined in the 
unlimited interval [0, $+\infty$), is a Rayleigh distribution, suppressed at large 
$x$ by a factor similar to that of a Fermi-Dirac (FD) distribution:

\begin{equation}\label{e:mstdistr2}
f(x) = K~\frac{x}{\sigma^2}~\exp \left\{ -\frac{(x - \mu)^2}{2\sigma^2} \right\} 
\cdot \frac{1}{\exp\left( \frac{x - c}{d} \right) +1}
\end{equation}

The parameters values were found by means of a numerical best fit and
the resulting formula is: 
\begin{equation}\label{e:mstdistr3}
f(x) = \frac{5}{3}~x~\exp \left\{ -\frac{(x + 0.3)^2}{2.16} \right\} \cdot 
\frac{1}{\exp\left( \frac{x - 1.81}{0.156} \right) +1}
\end{equation}
with a maximum error with respect to the data less than 2\%.
We computed the values of the mode, the median, the variance and other moments 
from this distribution, and found 0.892 and 0.952 respectively for the first two, 
a variance equal to 0.208, whereas the skewness and the kurtosis are 0.080 and 
2.439, respectively.

Another fitting formula can be obtained from Pearson distributions (Smart 1958, 
chap. 7), again suppressed at large $x$ values by a FD factor:
\begin{equation}\label{e:mstdistr1}
\begin{split}
f(x) &=  K~\left[\left(\frac{x}{a_1}\right)^{b a_1}\right]  
\left[\left(1+\frac{a_1}{a_2} - \frac{x}{a_2}\right)^{b a_2}\right]  \cdot \\
& \cdot \frac{1}{\exp\left( \frac{x - c}{d} \right) +1}
\end{split}
\end{equation}
where $K$ is a normalization factor, $a_1$ is the value of the mode, $b$ and 
$a_2$ are free parameters, $c$ is the cut-off scale.
Differently from Eq. \eqref{e:mstdistr2}, this distribution is defined in the 
finite interval $x \in$ $[0, a_1 + a_2]$. 
Considering that values of $x$ larger than 3.0 are extremely rare, we imposed 
the condition $a_1 + a_2 = 3.2$ and evaluated the remaining parameters. 
A very good fit was obtained for $a_1=0.91$, $b=1.25$, $c=1.8$, $d=0.18$ and 
the normalisation factor $K=0.7676$.

The edge distribution can be useful for the choice of the separation parameter 
$\Lambda_{\mathrm{c}}$. 
From Eq. \eqref{e:mstdistr3} and Figure \ref{fig:hist3},  we can see
that the choice of a low $X_{\mathrm{c}} =\Lambda_{\mathrm{c}}/\Lambda_{\mathrm{m}}$, 
for instance the value of 0.37, 
implies that about 90\% of edges will be eliminated, and the majority of remaining
clusters will have a number of nodes too small to satisfy the elimination criteria.  
A good choice is to use a value close to unity: we found from our simulations that
the best range for $X_{\mathrm{c}}$ is between 0.8 and 1.2, corresponding to
the cumulative probabilities of 0.384 and 0.683, respectively. 
In fact, although the probability to find an edge smaller than $\sim\Lambda_{\mathrm{m}}$ 
is still large, it is unlikely that a high number of these edges will belong to 
a single remaining cluster and they are therefore rejected by the subsequent filtering.  

\subsection{Distribution of the number of sub-trees for a given $\Lambda_{\mathrm{c}}$ 
in a random field}

As shown by Di Ges\`u and Sacco (1983), the expected total number of clusters 
obtained by cutting a random, 2-dimensional MST having $N_{\mathrm{tot}}$ points, 
at an edge length $\Lambda_{\mathrm{c}}$, is given by:
\begin{equation}
\bar{N} = 1 + (N_{\mathrm{tot}} - 1) \exp \left\{ - \pi  \Lambda_{\mathrm{c}}^{2} 
N_{\mathrm{tot}}/A \right\}
\end{equation}
where $ N_{\mathrm{tot}}/A$ is the density of nodes, that according Eq. 
\eqref{e:meanlength} is proportional to $1/\Lambda_{\mathrm{m}}^2$.
This is a monotonic decreasing function, and we verified with Monte Carlo 
simulations the consistency of this result.

\begin{figure}
\centering
\includegraphics[scale=0.3, angle=270]{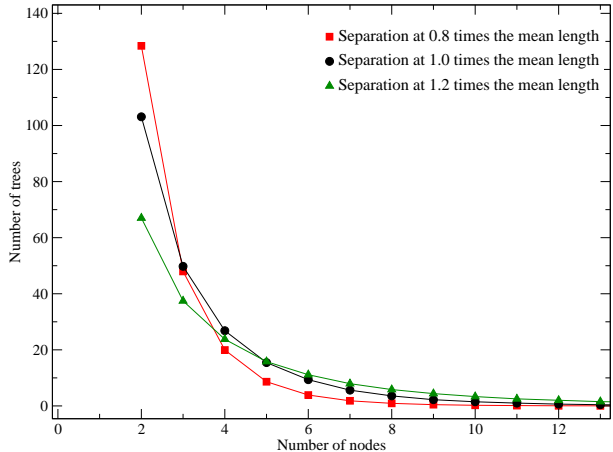}
\caption{Average number of sub-trees obtained separating a 1000 points random 
field at 0.8, 1.0 and 1.2 times the mean MST length.}\label{fig:ntrees}
\end{figure}

We used a different approach, directly based on the calculated mean edge length 
and considered another distribution, 
useful for selecting the best $N_{\mathrm{c}}$ parameter, 
that of the number of clusters as a function of the number of nodes after the 
application of a separation at the edge length $\Lambda_{\mathrm{c}}$. 
We computed several distributions in random fields via Monte Carlo simulations and 
found that they can be well described by an exponential function: 
\begin{equation}
T(N_{\mathrm{n}}) = F(X_{\mathrm{c}}) \cdot N_{\mathrm{tot}} 
\cdot e^{-\kappa(X_{\mathrm{c}}) N_{\mathrm{n}}}
\end{equation}
where $T(N_{\mathrm{n}})$ is the total number of sub-trees having $N_{\mathrm{n}}$ 
nodes each and $X_{\mathrm{c}}  = \Lambda_{\mathrm{c}}/\Lambda_{\mathrm{m}}$.
Some examples, corresponding to different choices of the cut length 
$\Lambda_{\mathrm{c}}$, are shown in Fig. \ref{fig:ntrees}.
We see how the mean number of big clusters decreases when the cut length becomes 
smaller than the mean MST edge length: that is explained by the fact that separating 
at smaller lengths (thus removing more edges from the MST) we tend to ``fragmentate'' 
the tree in more small pieces.

Considering the MST of a field with a total number of points $N_{\mathrm{tot}}$ 
and applying a separation at $\Lambda_{\mathrm{c}}$, we can establish an useful 
lower-limit for the elimination value $N_{\mathrm{c}}$ by comparing it with a 
random field with the same number of points and separated at the same length. 
A simple criterion is to choose the $N_{\mathrm{c}}$ value for which, in the 
corresponding random field, on average there is only one sub-tree with the same node 
number:
\begin{equation}\label{eq:ncut}
T(N_{\mathrm{n}}) \leq 1 \Leftrightarrow %
N_{\mathrm{c}} \geq N_{\mathrm{c}}^{1} = \frac{\ln(F(X_{\mathrm{c}}) \cdot 
N_{\mathrm{tot}})}{\kappa(X_{\mathrm{c}})}
\end{equation}
Another possibility is to use the value $N_{\mathrm{c}}^{\star}$ for which the 
expected number of residual casual clusters is less than unity, which is obtained 
by integrating the distribution of Eq.(5):
\begin{equation}\label{eq:ncutstar}
\int_{N_{\mathrm{c}}^{\star}}^{\infty} T(N_{\mathrm{n}})dN_{\mathrm{n}} 
\leq 1 \Leftrightarrow %
N_{\mathrm{c}}^{\star} \geq N_{\mathrm{c}}^{1} - \frac{\ln\kappa(X_{\mathrm{c}})}
{\kappa(X_{\mathrm{c}})}
\end{equation}
Note that $N_{\mathrm{c}}^{\star}$ is slightly greater than the value given by 
Eq.~\eqref{eq:ncut}.
Of course, the choice $N_{\mathrm{c}} > N_{\mathrm{c}}^{\star}$ would give a higher
confidence on the source detection, but the risk of eliminating true faint sources 
increases.  

The two functions $F(X_{\mathrm{c}})$ and $\kappa(X_{\mathrm{c}})$  are
characterized by a monotonic decreasing behaviour and are well described by
the following power laws:
\begin{equation}
F(X_{\mathrm{c}}) = 0.2\, X_{\mathrm{c}}^{-3.74}
\end{equation}
and
\begin{equation}
\kappa(X_{\mathrm{c}}) = 0.5\, X_{\mathrm{c}}^{-1.93}
\end{equation}
In a random field we have $F(X_{\mathrm{c}}) = 0.461,\, 0.200,\, 0.101$ and 
$\kappa(X_{\mathrm{c}}) = 0.77,\, 0.50,\, 0.35$ for $X_{\mathrm{c}} = 0.8,\, 1.0,\, 
1.2$, respectively.

\begin{figure}
\centering
\includegraphics[scale=0.3, angle=270]{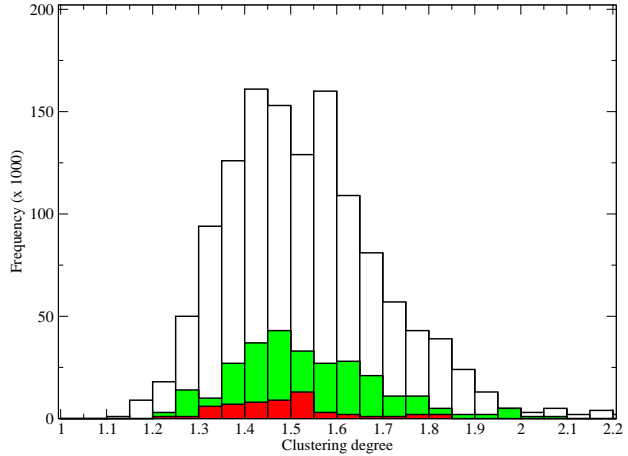}
\caption{Frequency distribution of the clustering degree for the residual 
clusters, for $N_\mathrm{c} = 12$ (white), 16 (green/light dark) and 20 (red/dark).}
\label{fig:hist4}
\end{figure}

\section{Clustering parameter and detection stability} \label{sect:bootstrap}

\subsection{Clustering parameter}
Once a list of candidate sources is found, it is useful to introduce some 
criteria to select, among the sub-trees remaining after the application of the 
filters, those corresponding to the best candidate sources and to reject clusters 
with a high chance to be randomly originated. 
 
A first parameter is the ``clustering degree'' that we define as 
$g = \Lambda_{\mathrm{m}}/\Lambda_{\mathrm{m,tree}}$, i.e.
the ratio between the mean edge length in the whole MST to the one 
of the edges in the sub-tree. 
The more clusterized is a sub-tree (then, the more likely is the candidate source 
to be true), the less will be its mean edge length $\Lambda_{\mathrm{m,tree}}$ 
and the bigger will be the value of the clustering degree $g$.

We tested how $g$ works investigating its distribution for clusters in a random  
field.
We generated 1000 fields of 1000 points each and evaluated $g$ for the remaining
clusters, after a separation at $\Lambda_{\mathrm{c}} = \Lambda_{\mathrm{m}}$ and 
elimination at $N_\mathrm{c} = N_\mathrm{c}^{\star} = 12$. 
In Fig. \ref{fig:hist4} we show the histogram of the resulting distribution of $g$ 
for the residual clusters, about one for each field as expected.   
It has an approximate Gaussian shape, 
although with asymmetric tails.
The maximum is around $g=1.5$ and the skewness has the low value of 
$\beta_{3}\simeq0.2$. 
An acceptable fit can be obtained by the same Pearson distribution used in
Eq. \eqref{e:mstdistr1}, without the exponential suppression factor. 
For comparison, the distributions of $g$ in which the elimination value is raised 
to  $N_\mathrm{c} = 16$ and 20 are also shown. 
It's evident that the mean clustering degree of the residual clusters is still 
around $g=1.5$, but the frequency of these clusters is much lower.
We can conclude that, 
in a ``true'' field with the same number of points and separated at the same length, 
clusters with $g > 1.7$ 
combined with a number of nodes sufficiently higher than $N_{\mathrm{c}}^{\star}$,
are good candidates to be genuine sources. 
For example, in our simulations, cutting respectively at $N_\mathrm{c} = 
12$, 16 and 20 we have frequencies of random clusters with $g > 1.7$ 
 of about 20\%, 4\% and 0.5\%, respectively.
A higher threshold value of $g$ would result in a safer rejection of spurious 
clusters, but in this case it is also possible to eliminate real weak sources.
A good choice can be reached by comparing the values of $g$ between the 
remaining clusters.

For the two clusters of Fig. 1 (lower-right panel) we have $g=2.21$ and $g=1.96$
for the ``strong'' and the ``faint'' source, respectively, while lowering the $N_{\mathrm{c}}$ value
below $N_{\mathrm{c}}^{\star} = 10$ spurious clusters also with $g>1.7$ appear.

\subsection{Bootstrap method and detection stability}
We can define another parameter to take into account that the position of individual 
events in $\gamma$-ray images does not coincide with the true incoming direction of 
photons, because the typical uncertainty due to the reconstruction of pair trajectories 
is of the order of a few degrees.
A $\gamma$-ray image, therefore, must be considered as a possible realisation 
of a large set of images of the ``true'' field in the sky. 
To take into account this effect and to verify if the detected clusters were 
produced by casual aggregation of events or can be considered associated with 
real sources, we introduced a ``bootstrap'' technique that 
can be used to improve the confidence on detections.
Starting from the original image, we produce a set of other possible images by 
generating an equal number of photons whose coordinates are randomly extracted 
with a probability density function approximating the instrumental PSF, 
and including the energy dependence.  
We then apply the MST algorithm to these bootstrapped fields, with the same filter 
selection as in the original one, and as output we obtain new lists of candidate 
sources to be compared with the original detections.
Those having positions within the PSF size are assumed to correspond to the same 
original source.
Candidate sources having a high detection frequency in the bootstrapped images correspond 
to rich and dense clusters and have a high probability to be real, whereas 
sources characterised by a small number of nodes and a low clustering degree $g$
are generally detected with a low frequency.
The ``detection stability'' parameter $s$ is then given by the ratio of the
number of detections inside a source circle to the total number of bootstrapped 
fields. 
Sometimes a single cluster is divided into a couple of smaller clusters inside
the source circle. 
In this cases, smaller sources are counted as a single detection to avoid that
$s$ can result higher than unity. 
From our simulations we found that one can consider a source detection 
as reliable if the corresponding cluster is
detected in at least one half of the bootstrapped fields, i.e. with $s \geq 0.5$.

\begin{figure*}
\centering
\includegraphics[scale=0.6]{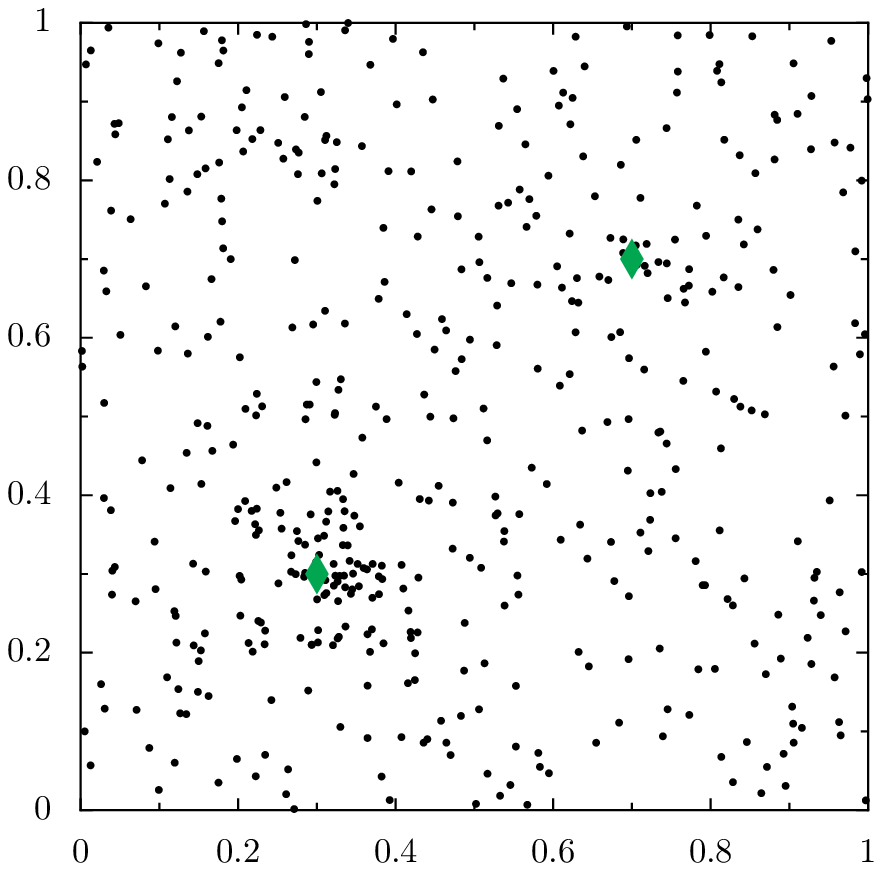} \includegraphics[scale=0.6]{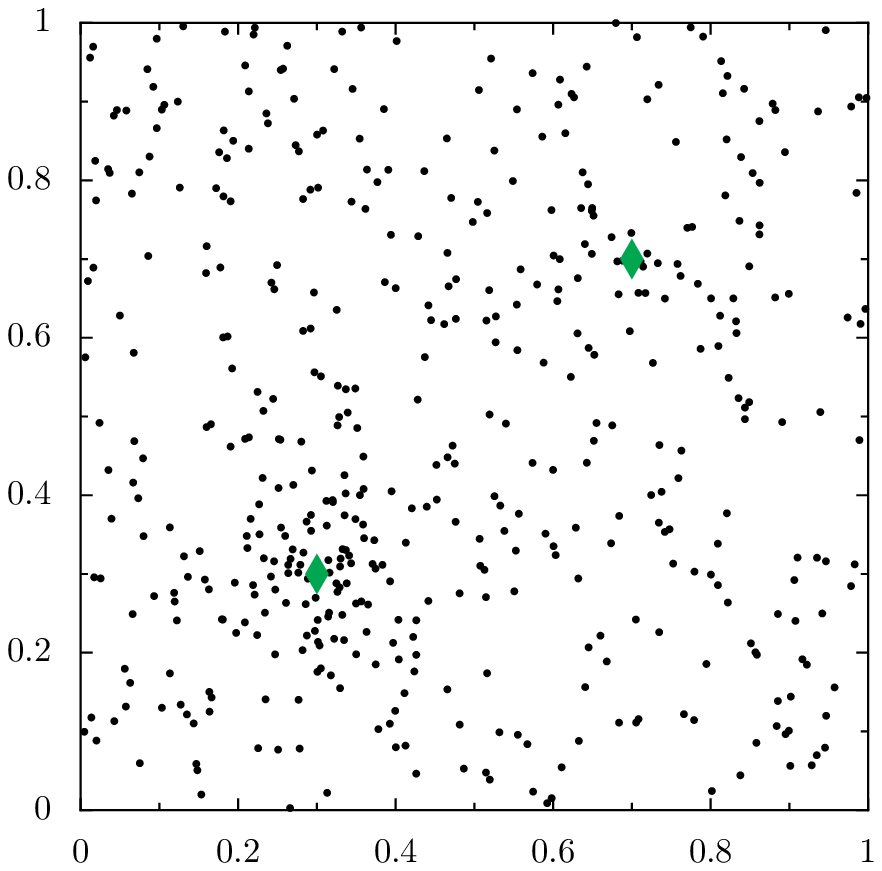} \\
\includegraphics[scale=0.6]{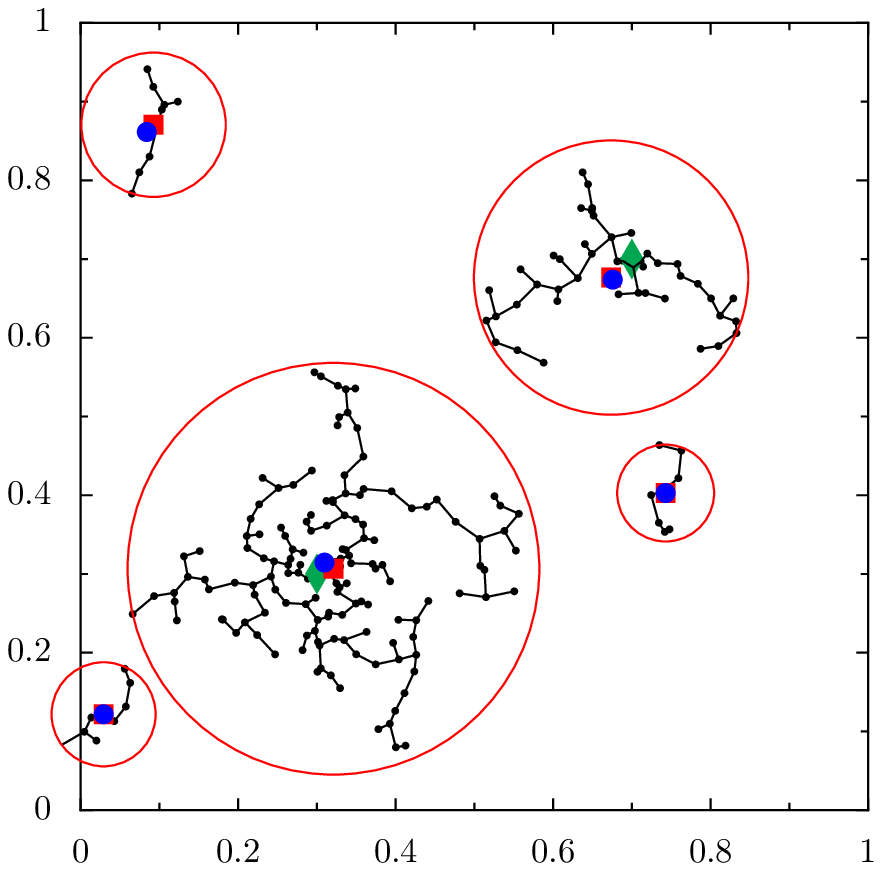} \includegraphics[scale=0.6]{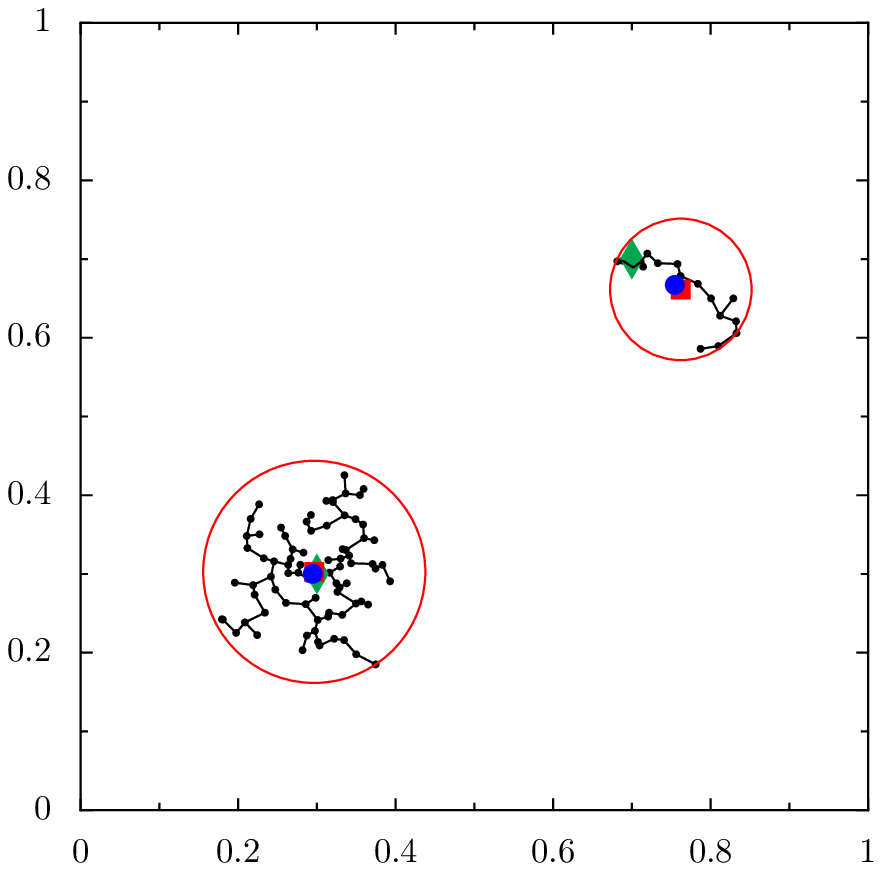}
\caption{\emph{Upper left}: The 500-point field of Fig. 1, with two simulated sources.
\emph{Upper right}: A bootstrap realisation of the same field, with $\sigma=0.1$.
\emph{Lower left}: Cluster selection after the applications of the filters $\Lambda_{\mathrm{c}}= 
1.3\,\Lambda_{\mathrm{m}}$ and  $N_{\mathrm{c}}= 7$ onto the bootstrapped field.
\emph{Lower right}: Cluster selection after  $\Lambda_{\mathrm{c}}= 
\Lambda_{\mathrm{m}}$ and  $N_{\mathrm{c}}= 10$ onto the bootstrapped field.
The meaning of symbols is the same as in Fig. 1.
}
\label{fig:bootexample}
\end{figure*}

An example is given in Fig. \ref{fig:bootexample},
 where we show a bootstrap of the Fig. 1 field. 
In the upper left panel the original field is shown, 
while the upper right panel is a bootstrapped field computed
using a probability density function equal to the one used to generate the simulated 
sources, i.e. a Gaussian with $\sigma = 0.1$. 
Note that the strong source is more or less unaffected by the redistribution of photons,
whereas some other small clusters appear, but they are rejected by further filtering.
In the lower panels the clusters remaining after the MST application (left: $\Lambda_{\mathrm{c}}= 
1.3\,\Lambda_{\mathrm{m}}$,  $N_{\mathrm{c}}= 7$, right: $\Lambda_{\mathrm{c}}= \Lambda_{\mathrm{m}}$,  $N_{\mathrm{c}}= 10$) onto this particular bootstrapped field: note the different shape and number of clusters with respect to the corresponding panels in Fig. 1.
With the generation of 100 bootstrap fields and the first selection of filters, we obtain a detection stability of $s=1$ for both sources.
If we choose the second set of filters the detection stability is $s=1$ and $s=0.55$
for the ``strong'' and the ``faint'' source, respectively.

The bootstrap method can also be used when the MST algorithm detect 
two very close clusters, with a separation between the centroids less than the PSF size,
an effect likely due to the presence of an edge just above the cutting
threshold.
In this case a large fraction of the bootstrapped fields has a single cluster  
at the position of these two sub-trees and we can conclude that the splitting
into two clusters was accidental and that they correspond to a unique source.

The statistical distribution of the expected values of $s$ in a random field 
cannot be computed because it depends upon the instrumental response functions 
used when bootstrapped coordinates are generated. 
An estimate of the threshold $s$ value to reject unstable clusters must be then 
obtained from a comparison of the resulting values.   
We also noticed in our simulations that the source position computed averaging
the centroids of the bootstrap replicas is frequently, although not always, closer
to the actual source location than that derived from the MST application to the
original field. 
Eventually, this method can be also used to refine the source coordinates.

\begin{figure*} 
\centering
\includegraphics[scale=0.35]{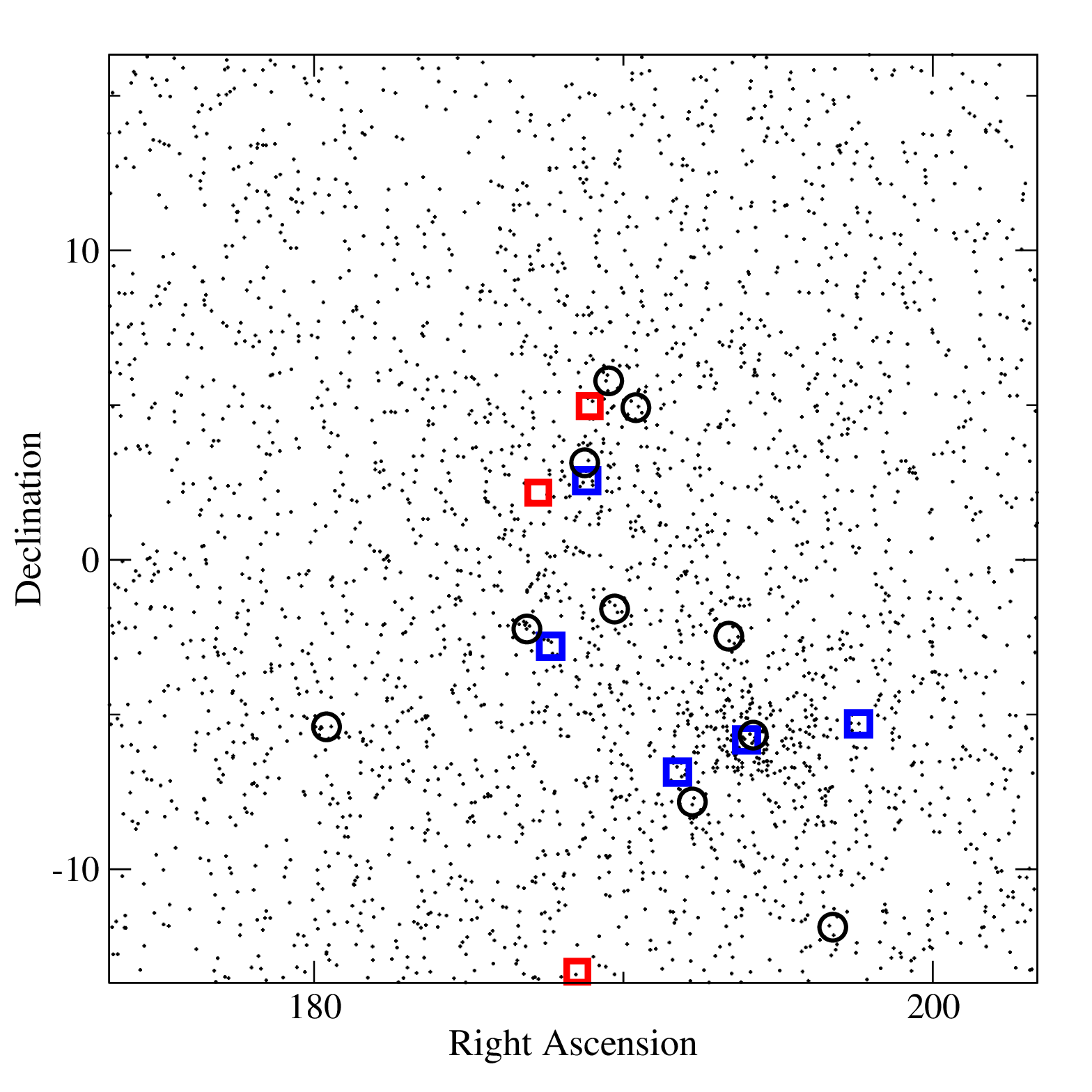}\includegraphics[scale=0.35]{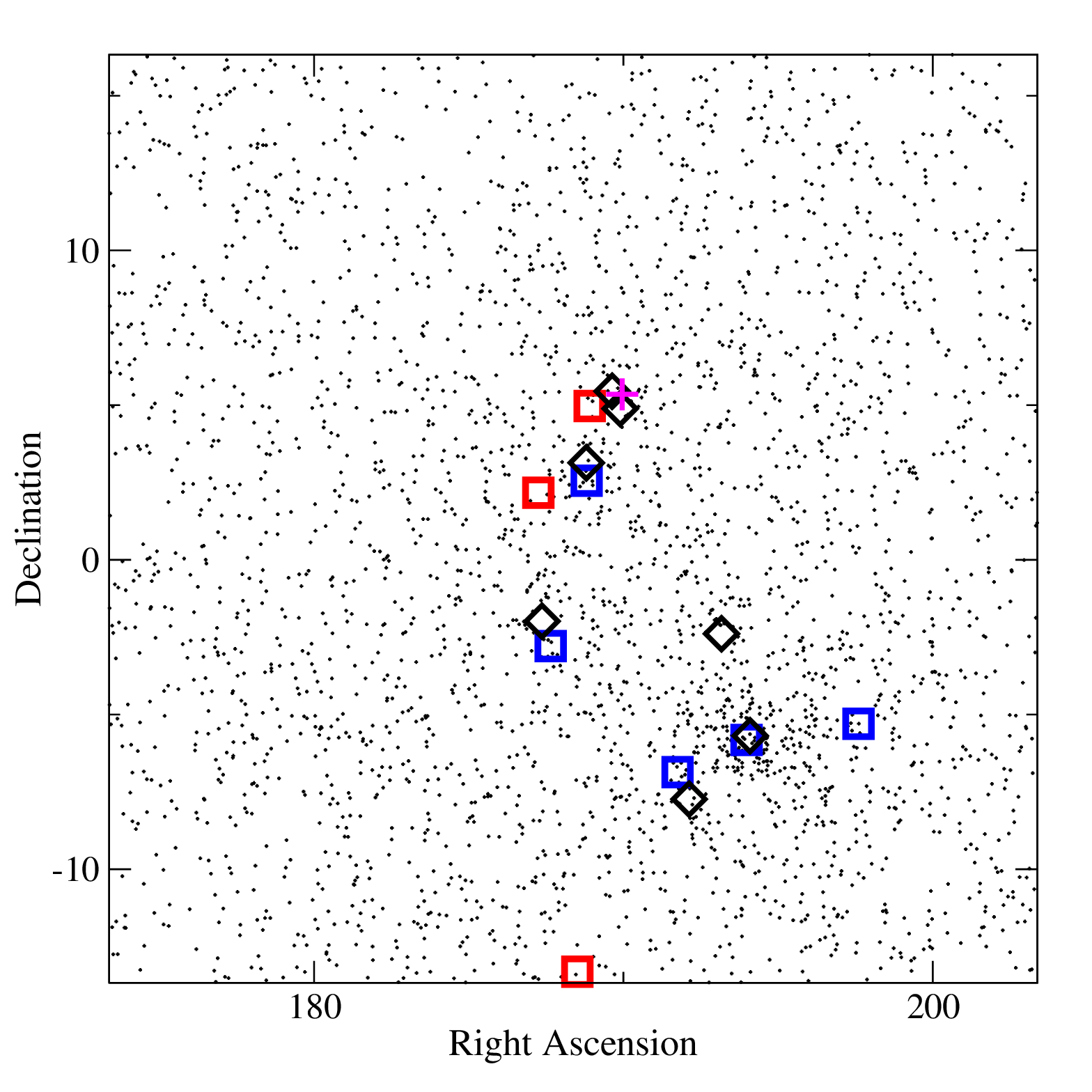}
\caption{\emph{Left:} 
EGRET-VP11.0 pointing, 30$^{\circ}$$\times$30$^{\circ}$\ square field, centered on 
RA 188.38, DEC +1.33. 
Blue/dark squares are the 3EG sources detected in this pointing, red/light dark squares the other 3EG 
sources within this area. Black circles are the MST-detected candidate sources.
\emph{Right:}
The black diamonds are the positions of candidate sources calculated via the bootstrap method. 
Only the sources with a clustering degree $g>1.70$ and with a bootstrap detection 
stability $s > 0.5$ are retained.
The cross here mark the mean coordinates of the two clusters that likely belong to the same
source, see text for discussion.
}\label{fig:egret_pnt110}
\end{figure*}

\section{Application of MST to EGRET fields}

To test the source detection capability of our implementation of the MST,
 we applied it to a real $\gamma$-ray image in which several sources were already found. 
Due to the simplicity of our algorithm (for example, we don't treat
the energy dependence of the point-spread function 
and the geometrical distortions of photon distribution in a flat projection),
we choose an high galactic latitude field in order to have a low, uniform background, 
with no strong intensity gradient.
Morover, this field lies around the celestial equator and projection effects
on photon coordinates are negligible, being smaller than a few percent.

Fig. \ref{fig:egret_pnt110} shows the central portion of the EGRET  Cycle 1 VP-11.0 
field for photon energies higher than 100 MeV\footnote{A standard energy-dependent 
cut on zenith angle has been applied to the original photon list, in order to remove 
Earth albedo $\gamma$-ray background (Esposito et al. 1997).}, 
observed between 03 and 17 October 1991 and comprising the
 quasars 3C~273, 3C~279 and other sources. 
In the left panel, blue squares mark the sources detected in this specific pointing 
and reported in the Third Egret Catalog (3EG, Hartman et al. 1999), while the red 
squares are other 3EG sources in field but not detected in this pointing (i.e. only 
upper limits on the flux are given in the catalog). 

A first application of MST, using the filtering parametrs
$\Lambda_{\mathrm{c}}= 0.9\, \Lambda_{\mathrm{m}}$ and $N_{\mathrm{c}}^{\star} = 12$,
gave the detection of 10 clusters, shown as black circles in the same figure. 
We then used the bootstrap method (see Sect. 4) sorting new photon directions with a  
Gaussian distribution centered at each original point and having a 
$\sigma = 2$$^{\circ}$. 
Note that $a$) the choice of a Gaussian distribution for the bootstrapped photons 
is a simplified and energy-averaged approximation of the instrumental PSF, and 
$b$) the use of a circle for the computation of $s$ is only a zeroth-order 
approximation of the actual photon distribution, that for real astronomical data 
would be rather an ellipse, due to geometrical projection effects.
Over 100 bootstrap fields were so produced, and only the seven candidate sources with a clustering 
degree $g>1.70$ and with a detection stability $s > 0.5$ were retained. 
The MST detected clusters satisfying these criteria are given in Fig. \ref{fig:egret_pnt110}
(right panel) and Table 1, where
we report the coordinates, number of nodes, the clustering degree $g$, 
bootstrap detection stability $s$, the 3EG counterpart 
and the possible identification based on the new catalogue of blazars Roma-BZCat 
(Massaro et~al. 2005). 

For this pointing, the 3EG catalogue reports five sources, while two other sources 
are detected in other pointings of the same field.
Five of these seven 3EG sources were also detected by the MST method 
and their angular distance from the catalog positions is $\sim 1$$^{\circ}$\ or less.
Four correspond to the 3EG sources detected in this pointing, 
while the fifth, 3EG~J1310--0517, which is reported as an unidentified and possibly 
confused source, is not detected (although there is a 12-node cluster that correspond 
to this source after the separation, thus just below the elimination value). 
We found also some additional clusters. 
Two of them (RA=189.52$^{\circ}$, $\delta$=5.78$^{\circ}$; RA=190.40$^{\circ}$, 
$\delta$=4.92$^{\circ}$) have the small separation of $\sim$1$^{\circ}$\ and lie 
close 3EG~J1236+0457, which was not detected in this pointing by Hartman et al. (1999).
The distance between their centroids is less than the PSF radius, so it is very 
likely that they belong to a single source: the sub-tree of this source was split into 
two sub-trees by removing one single edge having a length slight exceeding 
$\Lambda_{\mathrm{c}}$. 
This is confirmed by the fact that in about one half of the bootstrapped fields 
there is a single cluster near this position.
For this reason we can consider the two clusters as belonging to a unique
source located approximately at the mean position 
(RA=189.96$^{\circ}$, $\delta$=5.35$^{\circ}$).
This source is likely associated with the $z=1.762$ flat spectrum radio quasar 
BZQ~J1239+0443. 
We investigated whether in other EGRET pointings containing the same region of 
the sky 
this source is present, and detected it in VP-408.0 and VP-306.0. 
In the former pointing the source was also detected by Hartman et al. (1999), but not 
in the latter. 
Although, we didn't found it in VP-407.0 where it was reported by Hartman et al. (1999).
This discrepancy is to be attributed mainly to the faintness of these sources, 
which make the detection extremely sensitive to the actual source detection method 
used and its threshold values.
 
MST algorithm detected a cluster at the coordinates (RA=193.41$^{\circ}$, 
$\delta$=-2.47$^{\circ}$), 
which is not in the 3EG catalogue.
In particular, this cluster has MST parameters comparable to those of 3C~273 and
there is no reason to reject it.
We searched without success in the Roma-BZCat and in the NED database for possible 
counterparts and therefore it remains unidentified. 
Of course, the possibility that it must not be considered genuine and originated by
random clustering of events in the field cannot be excluded.

\begin{table*}
\caption{MST-detected clusters in EGRET pointing 110.0, with $g > 1.7$ and $s > 0.5$. 
For each candidate source are reported 
the celestial coordinates (Right Ascension and Declination, in degrees), the number 
of nodes of the relative cluster, the clustering degree $g$, the bootstrap detection 
stability $s$, the Third EGRET Catalog (3EG) counterpart, and the identification 
with known sources. 
}
\label{tab:egret_pnt110}
\centering
\begin{tabular}{ccccccc}
\hline
RA & DEC & $N_{\mathrm{n}}$ & $g$ & $s$ & 3EG counterpart & Identification \\
\hline
194.20 & -5.66 & 201 & 2.31 & 1.   & 3EG J1255--0549 & 3C 279 \\
193.41 & -2.47 & 21  & 1.88 & 1.   & -- &  -- \\
192.22 & -7.82 & 37  & 1.83 & 1.   & 3EG J1246--0651 &  BZB J1243--0613 \\ 
190.40 &  4.92(*) & 16  & 1.79 & 1.   & 3EG J1236+0457 & BZQ J1239+0443 \\
188.75 &  3.14 & 23  & 1.71 & 1.   & 3EG J1229+0210 & 3C 273 \\
189.52 &  5.78(*) & 17  & 1.71 & 1.   & 3EG J1236+0457 & BZQ J1239+0443 \\
186.88 & -2.23 & 15  & 2.35 & 0.92 & 3EG J1230--0247 & BZQ J1236+0224 \\
\hline
\end{tabular}
\\
(*)~These two clusters likely correspond to a unique source, located at about
(189.96, 5.35), \\ as indicated by the fact that their clusters are connected in
about half of boostrapped fields.
\end{table*}

\section{Discussion}

We presented an application of a Minimal Spanning Tree algorithm to the problem 
of source detection in $\gamma$-ray images.
This method does not involves in the computation the instrumental response functions
and works recognizing the regions of the sky where arrival directions of photons 
clusterize.
It has the advantages of a fast calculation but did not provide directly estimates of
the source flux.
We have shown that a MST based algorithm is a viable method to detect $\gamma$-ray 
sources both in simulated images and in real $\gamma$-ray observations 
of the EGRET experiment on board Compton-GRO.  
We proposed some tools to optimize the filtering parameters and to assess the reliability 
of source detections, like the clustering degree and the bootstrap detection stability. 
These tools are based on a study, although empirical, of the statistical properties 
of the Minimal Spanning Tree on random fields. 

The MST application to an EGRET field around the two famous $\gamma$-ray loud quasars 
3C~273 and 3C~279 found almost all the 3EG sources already detected in the same pointing
and confirmed the presence of another source, detected in a different pointing.
We consider this result a good indication that MST method is particularly efficient.
We found also evidence of a new possible source with a significance comparable to that 
of other well established sources.
We expect that future experiments with a better sensivity, like the LAT instrument on 
board GLAST, will confirm or disprove this finding.   

There are, however, several possible effects that make difficult the source detection 
and require even more attention when the MST method is used.
These problems can be divided into four main categories:
$i$) problems due to the presence of strong sources, 
$ii$) problems arising from energy spectra of the sources different from that of the 
background; moreover, different spectral indices between the sources
will result in different probabilities to be detected, due to the energy dependence of the PSF,
$iii$) problems originated by images with a non-homogeneous background,
$iv$) problems due to the geometrical distortions from the arriving celestial photons 
in projection onto the $\gamma$-ray telescope, 
that will result not necessarily in a circular shape to characterize proper cluster selections.
At present we have not developed a well established strategy to solve these problems 
and in the following we will briefly discuss some aspects useful for the understanding 
of results.

One or more strong sources in the field have various possible consequences.
A first relevant effect is that they are characterized by a high clustering degree
and consequently reduce the value of $\Lambda_{\mathrm{m}}$ with respect to the one
expected in the field if they were absent.
A value of $\Lambda_{\mathrm{c}}$ very close to $\Lambda_{\mathrm{m}}$ 
would here be good to detect strong sources but this selection criterion 
could miss other possible sources of lower flux.  
Another effect is the presence of possible ``satellites'' in the surroundings of a 
strong source, even closer than expected from the PSF, originated by cutting an edge 
whose length is just smaller than $\Lambda_{\mathrm{c}}$. 
For example, the cluster detected in the EGRET field (see Sect.~4) with no obvious counterpart,
is at a distance of about 3.3$^{\circ}$ from the strong radio quasar 3C~279, 
and therefore we cannot exclude that it could be a satellite of the latter.
Usually, the satellites do not have a high frequency in the bootstrap
fields.

The energy distribution of the photons also affects the source detection, because the
PSF of $\gamma$-ray telescopes changes with the energy becoming much narrower at
high energies.
This implies that sources with spectra harder than the background are better detected 
in high energy images because their clustering degree increases.
At variance, sources with soft spectra give more disperse clusters and cannot be easily
found.

Another class of problems is present when the background is markedly non-homogeneous,
as in the case where the field contain a portion of the galactic disc.
In this case, using an unique $\Lambda_{\mathrm{c}}$ in all the image would correspond 
to a long cutting in the dense region and to a short cutting in the region of low 
density with the consequence of missing real sources and producing more spurious clusters.    

A general approach to be used for $\gamma$-ray source detection is that of using several
methods, possibly based on different techniques, and to compare their results.
In this way it will be possible to reduce the number of spurious detections, because
of the different criteria and \emph{a priori} assumptions applied in the source 
recognition. 
Accordingly, MST method can be used to obtain a quick list of photon clusterization 
regions, that could correspond to possible sources, to be studied indipendently 
with other methods.

There are other clustering algorithms that can be applied to $\gamma$-ray source
detection, like the Voronoi tessellation (Icke \& van de Weygaert 1987, Aurenhammer 1991). 
In particular, this method is based on the construction of its dual graph, the
Delaunay triangulation, of which MST is a subset.
We think, therefore, that at least in principle, they would provide similar result and
that a combined figure of merit for source detection should be defined.  
 
Here we discussed gamma-ray astronomy as a prime candidate for the application of 
MST method, 
but it could be even better applicable to the study of data clusterization 
in ultra-high energy cosmic rays (UHECR) and hemispherical neutrino experiments, that 
are characterized to the absence of structured background.
We think also that it will be possible to extend MST to higher dimensional spaces introducing
time and energy as additional dimensions.
Basically there are two approaches: $i$) to search for clusters in separate, 
dimensionally homogeneous subspaces, and then to search for the intersection of the 
detected clusters
and $ii$) to define a new metric for the tree edges that combine together the various
dimensions in a suitable way for the MST computation.
Preliminary numerical attempts based on the second approach, with energy as third 
coordinate, seem to be very promising to identify sources having spectra different 
from that of the background.  
Another possible 3-dimensional generalization is to take into account also the time, 
thus searching for variable or stable sources.
We will discuss a possible application of such a generalized MST in a subsequent work.

\section*{Acknowledgments}
We are grateful to Cettina Maccarone, Bruno Sacco, Paolo Giommi and Gino Tosti
 for useful discussions.
We also thank the anonymous referee for insightful comments.
This work is partially supported by ASI-INAF funds for the scientific programs
with GLAST.

\label{lastpage}
\end{document}